# Plasmon modes in graphene-GaAs heterostructures

Nguyen Van Men[1,2] and Nguyen Quoc Khanh[2]

[1]University of An Giang, 18-Ung Van Khiem Street, Long Xuyen, An Giang, Vietnam
[2]University of Science - VNUHCM, 227-Nguyen Van Cu Street, 5th District, Ho Chi Minh City, Viet Nam

___________________________________________________________________________

**Abstract**

We investigate the plasmon dispersion relation and damping rate of collective excitations in a double-layer system consisting of bilayer graphene and GaAs quantum well, separated by a distance $d$, at zero temperature with no interlayer tunneling. We use the random-phase-approximation dielectric function and take into account the nonhomogeneity of the dielectric background of the system. We show that the plasmon frequencies and damping rates depend considerably on interlayer correlation parameters, electron densities and dielectric constants of the contacting media.

___________________________________________________________________________

## 1. Introduction

Bilayer graphene (BLG), a special two-dimensional electron gas (2DEG) system, has attracted a great deal of attention in recent years because of its unique electronic properties [1-3]. A simple model for BLG single-particle spectrum is a pair of chiral parabolic electron and hole bands touching each other at the Dirac point [4]. The chiral electron wave function and parabolic dispersion lead to many properties significantly different from those of monolayer graphene (MLG) and ordinary 2DEG systems [5-6].

Plasmon excitations in many-electron systems have been studied long time ago and have been used to create plasmonic devices [7]. It was revealed that plasmon modes of graphene possess special properties [8] and the plasmon dispersion relation in double-layer systems differs significantly from the single layer one [9-14]. Plasmons in double-layer structures formed by 2DEG, MLG or BLG sheets of different type, separated by a dielectric film, may have more interesting properties because the carrier densities of 2DEG-graphene structure might be significantly different in both layers and the MLG-BLG structure is the massless-massive system [15-18]. Although double-layer systems are, in general, immersed in a nonhomogeneous three layered medium with background dielectric constants differing substantially from each others, many authors have used an average spatially independent dielectric permittivity to describe the plasmon energy dispersions of these systems [12-14]. It was shown that such description is inadequate and the nonhomogeneity of the dielectric background has been taken into account using the Poisson equation for the electrostatic problem in three-layer dielectric medium [18-19]. The authors of Refs. [15] and [16] have calculated the frequency of optical and acoustic plasmon modes of double-layer structures consisting of MLG and very thin 2DEG sheet at zero temperature. Principi and coworkers [15] have shown that long-range Coulomb interactions between massive electrons and massless Dirac fermions lead to a new set of optical and acoustic intra-subband plasmons. To our knowledge, up to now no similar calculations have been done for BLG-2DEG systems, although collective excitations of such chiral-nonchiral double-layer structures may have interesting properties.

Because of above reasons, in this paper, we consider a double-layer system consisting of doped BLG and GaAs quantum well, separated by a spacer of width $d$ assuming that 2DEGs in BLG and GaAs are electrically isolated [20]. We investigate the plasmon dispersion relation and damping rate at zero temperature using the random – phase – approximation (RPA) dielectric function, taking into account the nonhomogeneity of the dielectric background of the system.

## 2. Theory

We consider the double-layer system consisting of a BLG flake placed onto modulation-doped GaAs/AlGaAs heterostucture hosting a 2DEG, with the effective mass $m^*$, in the GaAs quantum well as shown in Fig. 1.

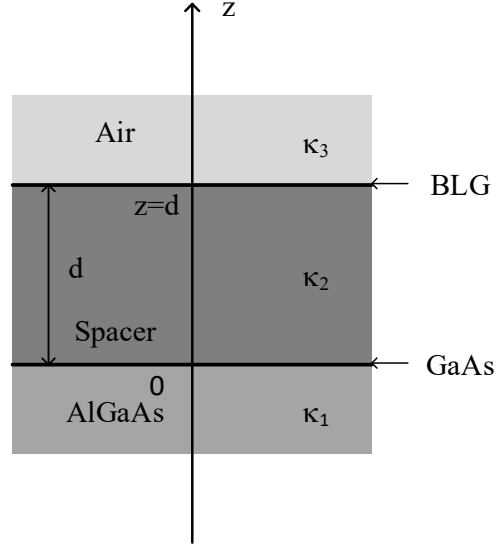

Fig. 1. A BLG-2DEG double-layer system immersed in a three layered dielectric medium with the background dielectric constants $\kappa_1$, $\kappa_2$ and $\kappa_3$.

The plasmon dispersion relation of an electronic system can be obtained from the zeroes of dynamical dielectric function [9-14]

$$\varepsilon(q, \omega_p - i\gamma) = 0 \qquad (1)$$

where $\omega_p$ is the plasmon frequency at a given wave-vector $q$ and $\gamma$ is the damping rate of plasma oscillations. In case of weak damping ($\gamma \ll \omega_p$), the plasmon dispersion and decay rate are determined from the following equations [11-12]

$$\operatorname{Re}\varepsilon(q, \omega_p) = 0 \qquad (2)$$

and

$$\gamma = \operatorname{Im}\varepsilon(q, \omega_p)\left(\frac{\partial \operatorname{Re}\varepsilon(q,\omega)}{\partial \omega}\bigg|_{\omega=\omega_p}\right)^{-1}. \qquad (3)$$

In the RPA, the dynamical dielectric function of BLG-2DEG double-layer system has the form [12-13, 19]

$$\varepsilon_{\text{2DEG-BLG}}(q,\omega) = \left[1 + U_{\text{2DEG}}(q)\Pi_{\text{2DEG}}(q,\omega)\right] \times$$
$$\times \left[1 + U_{\text{BLG}}(q)\Pi_{\text{BLG}}(q,\omega)\right] - [U_{\text{BLG-2DEG}}(q)]^2 \Pi_{\text{2DEG}}(q,\omega)\Pi_{\text{BLG}}(q,\omega) \qquad (4)$$

where $\Pi_{\text{2DEG}}(q,\omega)$ ($\Pi_{\text{BLG}}(q,\omega)$) is the zero-temperature non-interacting density-density response function of the 2DEG (BLG) given in [11, 21] ([6]). $U_{\text{2DEG/BLG}}(q)$ and $U_{\text{BLG-2DEG}}(q)$ are the intra- and inter-layer bare Coulomb interactions in momentum space [18],

$$U_{2DEG/BLG}(q) = \frac{4\pi e^2}{q} f_{2DEG/BLG}(qd) , \qquad (5)$$

$$U_{BLG-2DEG}(q) = \frac{8\pi e^2}{q} f_{BLG-2DEG}(qd) \qquad (6)$$

with

$$f_{BLG-2DEG}(x) = \frac{\kappa_2 e^x}{(\kappa_1 - \kappa_2)(\kappa_2 - \kappa_3) + e^{2x}(\kappa_1 + \kappa_2)(\kappa_2 + \kappa_3)} , \qquad (7)$$

$$f_{BLG}(x) = \frac{(\kappa_2 - \kappa_1) + (\kappa_1 + \kappa_2)e^{2x}}{(\kappa_1 - \kappa_2)(\kappa_2 - \kappa_3) + e^{2x}(\kappa_1 + \kappa_2)(\kappa_2 + \kappa_3)} , \qquad (8)$$

$$f_{2DEG}(x) = \frac{(\kappa_2 - \kappa_3) + (\kappa_3 + \kappa_2)e^{2x}}{(\kappa_1 - \kappa_2)(\kappa_2 - \kappa_3) + e^{2x}(\kappa_1 + \kappa_2)(\kappa_2 + \kappa_3)} . \qquad (9)$$

## 3. Numerical results

In this section, we calculate the frequency and damping rate of plasmon oscillations in BLG-2DEG double-layer system with $\kappa_1 = \kappa_{AlGaAs} = 12.9$, $\kappa_3 = \kappa_{air} = 1.0$ and $m^* = 0.067 m_0$ where $m_0$ is the vacuum mass of the electron at zero temperatures for several values of interlayer separation $d$, dielectric constant $\kappa_2$, graphene density $n_g$ and conventional 2DEG density $n_{2DEG}$. We show that Eq. (2) admits two solutions. The higher (lower) frequency solution corresponds to in-phase (out-of-phase) oscillations of densities in the two layers. These two branches of collective excitations of double-layer systems are known as optical and acoustic plasmon modes. In the following we denote the Fermi energy and Fermi wave-vector of BLG by $E_F$ and $k_F$, respectively.

In Fig. 2(a) we show the plasmon dispersion of BLG-2DEG double layer for $n_g = n_{2DEG} = 10^{12} \text{cm}^{-2}$, $d = 100 nm$ and $\kappa_2 = \kappa_{SiO_2} = 3.9$. The optical and acoustic plasmon dispersion curves of BLG-2DEG and 2DEG-2DEG systems for $n_g = n_{2DEG} = 10^{11} \text{cm}^{-2}$ with the same $d$ and $\kappa_2$ as in Fig. 2(a) are shown in Fig. 2(b) for a comparison. The solid (dashed) line is the optical (acoustic) plasmon dispersion and the dashed-dotted lines show the single particle excitations (SPE) boundaries of BLG and 2DEG. It is seen from Fig. 2(a) that the optical (acoustic) plasmon dispersion of BLG-2DEG system touches the edge of the continuum of BLG (2DEG) at a critical wave-vector. This behavior of optical (acoustic) plasmon in BLG-2DEG system is similar to that in BLG (2DEG) [6,9,11]. The Fig. 2(b) shows that the long wavelength optical plasmon of BLG-2DEG system has a $\sqrt{q}$ dispersion as that of 2DEG-2DEG double layer and is completely undamped [6, 15, 22]. In the region of larger wave-vectors the later is much higher than the former due to the effect of the chirality. Similar behavior has been found for single-layer systems [6, 22].

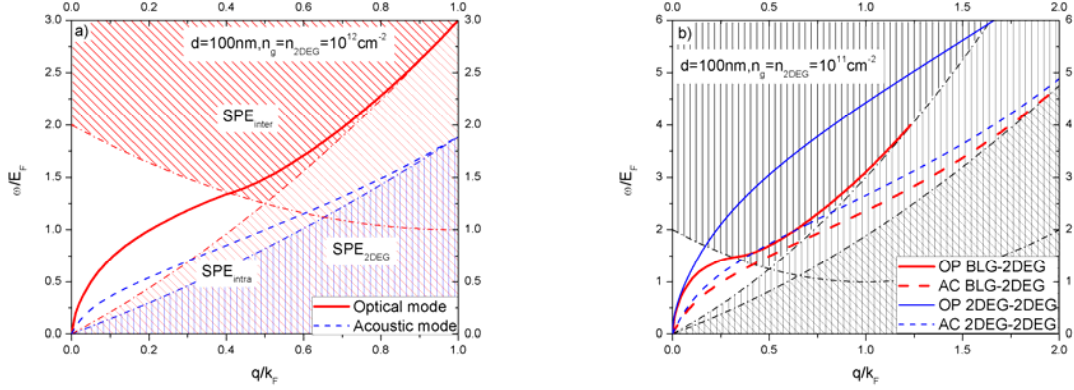

Fig. 2. Plasmon dispersion of BLG-2DEG and 2DEG-2DEG systems for $d=100nm$ and $\kappa_2 = \kappa_{SiO_2} = 3.9$. The two figures are for different densities (a) $n_g = n_{2DEG} = 10^{12} cm^{-2}$ and (b) $n_g = n_{2DEG} = 10^{11} cm^{-2}$. The dashed-dotted lines show the SPE boundaries of BLG and 2DEG (color online).

To see the effect of the interlayer distance, we show in Fig. 3 the plasmon dispersion of BLG-2DEG system with $\kappa_2 = \kappa_{SiO_2} = 3.9$, $d=50nm$ and 500nm for (a) $n_g = n_{2DEG} = 10^{12} cm^{-2}$ and (b) $n_g = n_{2DEG} = 10^{11} cm^{-2}$. We observe that both optical (solid curves) and acoustic (dashed curves) plasmon frequencies increase as the interlayer distance increases. The dependence on the interlayer distance is remarkable only at long wavelength limit.

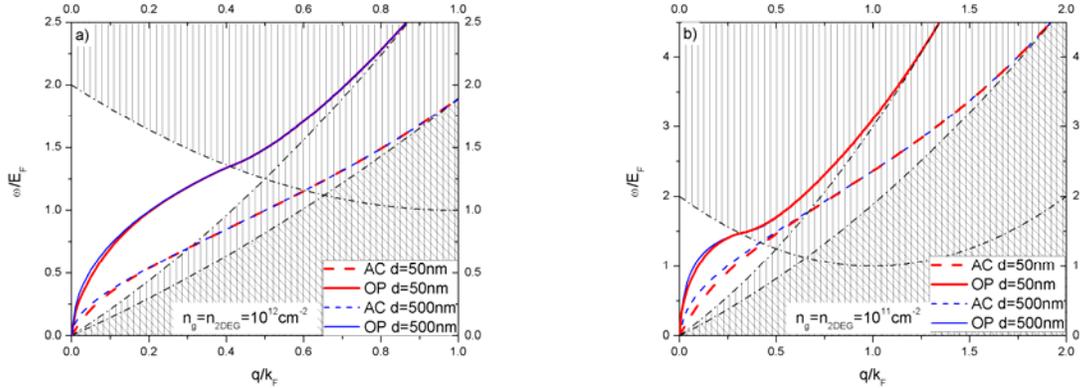

Fig. 3. Plasmon dispersion of BLG-2DEG system with $\kappa_2 = \kappa_{SiO_2} = 3.9$ for $d=50nm$ and 500nm. The two figures are for different densities (a) $n_g = n_{2DEG} = 10^{12} cm^{-2}$ and (b) $n_g = n_{2DEG} = 10^{11} cm^{-2}$. The dashed-dotted lines show the SPE boundaries of BLG and 2DEG (color online).

In order to understand the effect of different electron densities, we show in Fig. 4 plasmon mode dispersions of BLG-2DEG system with $\kappa_2 = \kappa_{SiO_2} = 3.9$ and $d=50nm$ for two cases: (a) $n_{2DEG} = 10^{12} cm^{-2}$, $n_g = n_{2DEG}$ and $2n_{2DEG}$, and (b) $n_g = 10^{12} cm^{-2}$, $n_{2DEG} = n_g$ and $2n_g$. The Fig. 4 indicates that the acoustic plasmon frequency (dashed curves) depends considerably on density ratio $n_g / n_{2DEG}$, especially for large wave-vectors. Both acoustic and optical frequencies decrease with increasing $n_g$, whereas the acoustic frequency increases significantly while the optical one decrease slightly with the decrease of $n_{2DEG}$.

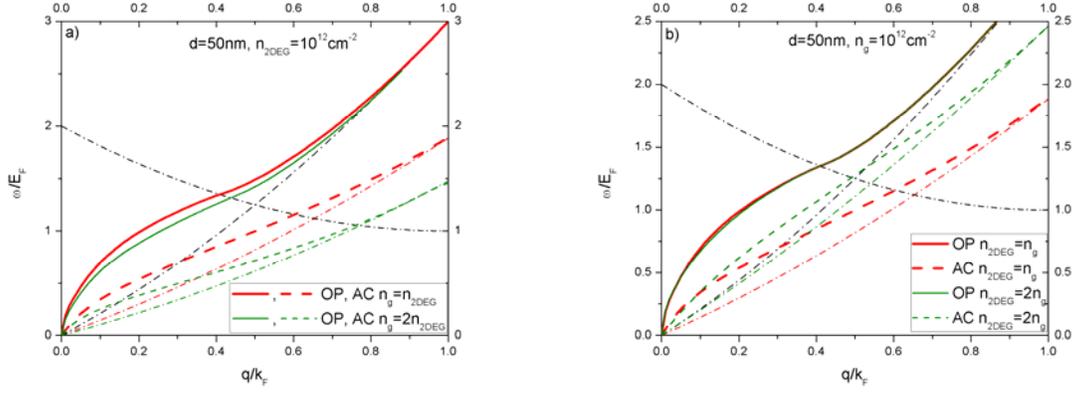

Fig 4. Optical (solid curves) and acoustic (dashed curves) plasmon modes of BLG-2DEG system for $\kappa_2 = \kappa_{SiO_2} = 3.9$, $d = 50 nm$. The two figures are for different densities (a) $n_{2DEG} = 10^{12} cm^{-2}$, $n_g = n_{2DEG}$ and $2n_{2DEG}$, and (b) $n_g = 10^{12} cm^{-2}$, $n_{2DEG} = n_g$ and $2n_g$ (color online).

In order to see the importance of nonhomogenous dielectric background, we show in Fig. 5 the plasmon dispersions calculated for nonhomogenous dielectric background with $\kappa_2 = \kappa_{SiO_2} = 3.9$ (solid curves) and for homogenous dielectric background with an average permittivity $\bar{\kappa} = (\kappa_1 + \kappa_3)/2 = 6.95$ (dashed curves) [19]. It is seen from the Fig. 5(a) (for $d = 50 nm$) and Fig. 5(b) (for $d = 500 nm$) that the nonhomogeneity of the background dielectric environment increases (decreases) the optical (acoustic) plasmon frequency. The effect of inhomogeneity of the dielectric background on the energy of optical plasmon modes is more pronounced than on that of acoustic ones. This behavior depends very weakly on the interlayer distance $d$.

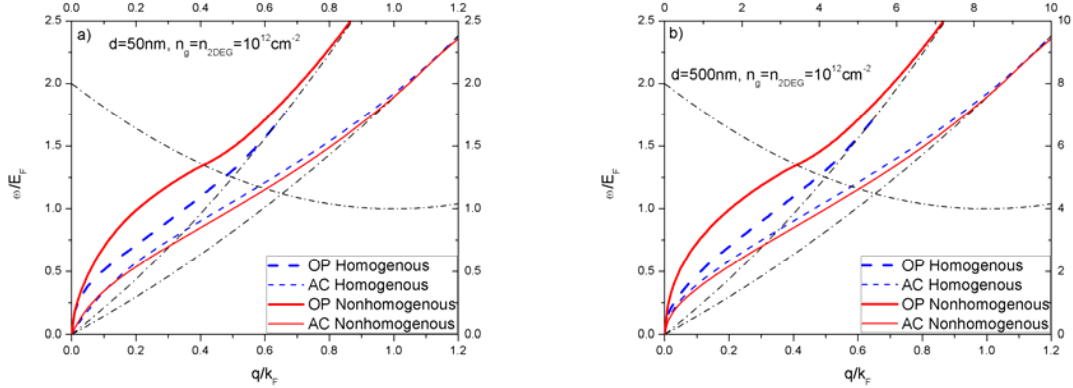

Fig. 5. Optical (upper curves) and acoustic (lower curves) plasmon modes calculated for nonhomogenous dielectric background with $\kappa_2 = \kappa_{SiO_2} = 3.9$ (solid curves) and for homogenous dielectric background with an average permittivity $\bar{\kappa} = (\kappa_1 + \kappa_3)/2 = 6.95$ (dashed curves) (color online).

The effects of the spacer on plasmon dispersions are shown in Fig. 6 for $\kappa_2 = \kappa_{SiO_2} = 3.9$ (thick curves) and $\kappa_2 = \kappa_{Al_2O_3} = 9.1$ (thin curves) [19]. Optical (solid curves) and acoustic (dashed curves) plasmon modes are calculated with $n_g = n_{2DEG} = 10^{12} cm^{-2}$ for a) $d = 50 nm$ and b) $d = 500 nm$. We observe that the energy of both optical and acoustic branches decreases when the dielectric constant of spacer increases. Again, this behavior is almost independent of the spacer width $d$.

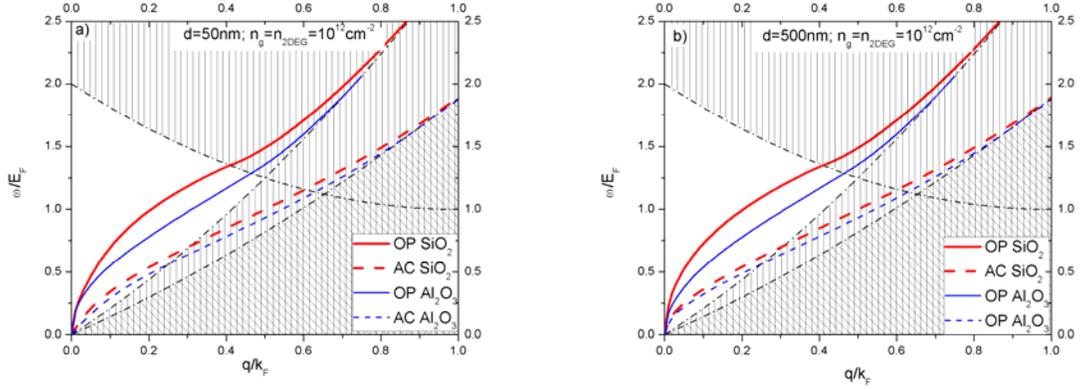

Fig. 6. Optical (solid curves) and acoustic (dashed curves) plasmon modes calculated for SiO$_2$ (thick curves) and Al$_2$O$_3$ (thin curves) spacer with $n_g = n_{2DEG} = 10^{12}\,\text{cm}^{-2}$ for a) $d = 50nm$ and b) $d = 500nm$ (color online).

Finally, we show in Fig. 7 the damping rate of optical and acoustic modes of BLG-2DEG double layer for SiO$_2$ spacer, $d = 100nm$ and $n_g = n_{2DEG} = 10^{12}\,\text{cm}^{-2}$. From Figs. 7 and 2(a), it is found that, the damping of optical mode increases as the dispersion goes into the BLG interband SPE region and then vanishes once the dispersion merges with the intraband continuum. The acoustic mode gets over damped when the plasmon dispersion touches the edge of the 2DEG continuum. We have also found that the damping rate of both plasmon branches depends considerably on density ratio and dielectric constant of spacer.

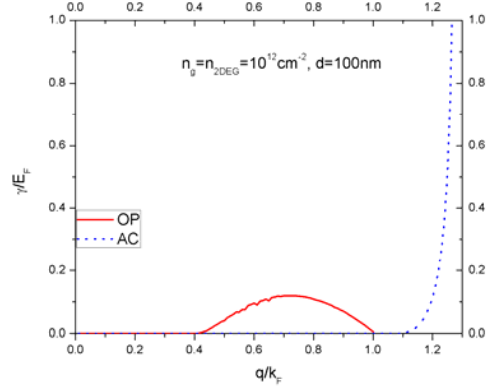

Fig. 7. Damping rate of optical and acoustic modes for SiO$_2$ spacer, $d = 100nm$ and $n_g = n_{2DEG} = 10^{12}\,\text{cm}^{-2}$ (color online).

## 4. Conclusion

In summary, we calculate for the first time the frequency and damping rate of plasmon oscillations in BLG-2DEG double layer at zero temperature using the RPA dielectric function, taking into account the nonhomogeneity of the dielectric background of the system. We show that the properties of the optical and acoustic modes depend considerably on the spacer width, electron densities and dielectric constants of the contacting media. It is found that, for system parameters used in this work, the frequency of both plasmon branches increases as interlayer distance increases and the dielectric nonhomogeneity increases (decreases) the optical (acoustic) plasmon frequency. We find that the energy of optical and acoustic modes decreases when the dielectric constant of spacer increases and the damping rate of both plasmon branches depends considerably on the density ratio and dielectric constant of spacer.

*Acknowledgement*

This research is funded by Vietnam National Foundation for Science and Technology Development (NAFOSTED) under Grant number 103.01-2017.23.